\documentclass[onecolumn,aps,letterpaper]{revtex4}
\usepackage{graphicx}

\newcommand{\BABARPubYear}    {08}

\newcommand{\BABARConfNumber} {020}
\newcommand{\SLACPubNumber} {13480}
\input pubboard/babarsym
 
\long\def\inst#1{\par\nobreak\kern 4pt\nobreak
    {\it #1}\par\vskip 10pt plus 3pt minus 3pt}

\begin{document}
{\thispagestyle{empty}

\begin{flushright}
\babar-CONF-\BABARPubYear/\BABARConfNumber \\
%\babar-PUB-\BABARPubYear/\BABARPubNumber \\
SLAC-PUB-\SLACPubNumber \\
%hep-ex/\LANLNumber \\
December 2008 \\
\end{flushright}

\par\vskip 5cm

\begin{center}
\Large \bf Search for the Lepton-Flavor Violating Decays \\
\Large \bf $\Upsilon(3S)\rightarrow e^{\pm}\tau^{\mp}$ and $\Upsilon(3S)\rightarrow\mu^{\pm}\tau^{\mp}$
\end{center}
\bigskip

\begin{center}
\large The \babar\ Collaboration\\
\mbox{ }\\
December 5, 2008
\end{center}
\bigskip \bigskip

\begin{center}
\large \bf Abstract
\end{center}
%%%%%%%%%%%%%%%%%%%%%%%%%%%%%%%%%%%%%%%%%%%%%%%%%%%%%%%%%%%%%%%%%%%%%%%%

Charged lepton-flavor violating processes are extremely rare in the Standard Model, but they are predicted to occur in 
several beyond-the-Standard Model theories, including Supersymmetry or models with leptoquarks or compositeness. We present
a search for such processes in a sample of $117\times10^6$ $\Upsilon(3S)$ decays recorded with the \babar\ detector. We 
place upper limits on the branching fractions $BF(\Upsilon(3S)\rightarrow e^{\pm}\tau^{\mp}) < 5.0 \times 10^{-6}$ and 
$BF(\Upsilon(3S) \rightarrow \mu^{\pm}\tau^{\mp}) < 4.1 \times 10^{-6}$ at 90\% confidence level. These results are used 
to place lower limits on the mass scale of beyond-the-Standard Model physics contributing to lepton-flavor violating decays
of the $\Upsilon(3S)$.

\vfill
\begin{center}

Submitted to the 6$^{th}$ International Workshop on Heavy Quarkonia \\
2-5 December 2008, Nara, Japan

\end{center}

\vspace{1.0cm}
\begin{center}
{\em SLAC National Accelerator Laboratory, Stanford University, 
Stanford, CA 94309} \\ \vspace{0.1cm}\hrule\vspace{0.1cm}
Work supported in part by Department of Energy contract DE-AC02-76SF00515.
\end{center}

\newpage
}

%% author list as of 03-Nov-2008 (481 authors)
%
\author{B.~Aubert}
\author{M.~Bona}
\author{Y.~Karyotakis}
\author{J.~P.~Lees}
\author{V.~Poireau}
\author{E.~Prencipe}
\author{X.~Prudent}
\author{V.~Tisserand}
\affiliation{Laboratoire de Physique des Particules, IN2P3/CNRS et Universit\'e de Savoie, F-74941 Annecy-Le-Vieux, France }
\author{J.~Garra~Tico}
\author{E.~Grauges}
\affiliation{Universitat de Barcelona, Facultat de Fisica, Departament ECM, E-08028 Barcelona, Spain }
% ...
\collaboration{The \babar\ Collaboration}
\noaffiliation

\maketitle 

\newpage

% The body of the paper starts here
\section{INTRODUCTION}
\label{sec:Introduction}

In the Standard Model, processes involving charged lepton-flavor violation (CLFV) are unobservable since they are suppressed by the ratio 
$(\Delta(m_{\nu}^2)/M_W^2)^2 < 10^{-48}$ \cite{ref:feinberg,ref:bilenky}. Here $\Delta(m_{\nu}^2)$ is the difference between the squared masses of 
neutrinos of different flavor and $M_W$ is the charged weak vector boson mass.  Many beyond-the-Standard Model (BSM) scenarios predict observable rates
for these processes \cite{ref:pati,ref:georgi}, which may lead to striking experimental signatures and provide unambiguous evidence for new physics. 
There have been considerable efforts both in experimental searches for CLFV decays of the $\mu$ and $\tau$ and theoretical predictions for their 
branching fractions in various BSM scenarios, but CLFV in the $\Upsilon$ sector remains relatively unexplored.  If new particles contributing to CLFV couple
to $b$ quarks, such processes may be observable in decays of the $\Upsilon$. Unitarity-based considerations
allow relations to be derived between CLFV decay rates of the $\mu$ and $\tau$ and corresponding
CLFV decay rates of the $\Upsilon$ \cite{ref:nussinov}. In particular, the current experimental constraint
$BF(\tau \rightarrow \ell\ell'\bar{\ell'})<(2-4) \times 10^{-8}$ \cite{ref:babarlfv,ref:bellelfv}, in which $\ell$ and $\ell'$ are charged
leptons of either the same or different flavor, implies
$BF(\Upsilon(3S) \rightarrow \ell^{\pm}\tau^{\mp})<(3-6) \times 10^{-3}$. A decay rate of this magnitude would
result in $O(10^{5})$ $\Upsilon(3S) \rightarrow \ell^{\pm}\tau^{\mp}$ decays in our dataset and would be easily
observable. If the new physics contributing to CLFV is in the Higgs sector, it would
preferentially couple to heavy quark flavors, further motivating the search for CLFV in
the bottomonium sector.

The rates for the CLFV decays $\Upsilon(4S) \rightarrow \ell^{\pm}\tau^{\mp}$ are too small to be observed,
since even $\Upsilon(4S) \rightarrow \tau^{+}\tau^{-}$ has not yet been observed. However, the branching fractions for 
rare decays of the narrow  $\Upsilon(3S)$ resonance are enhanced roughly 
by $\Gamma_{\Upsilon(4S)} / \Gamma_{\Upsilon(3S)} \approx 10^{3}$ with respect to those of the $\Upsilon(4S)$,
dramatically increasing the sensitivity to rare processes. In this analysis we search for the CLFV decays $\Upsilon(3S) \rightarrow \ell^{\pm} \tau^{\mp}$  $(\ell=e,\mu)$ 
using data collected with the \babar\ detector. The prior constraints on CLFV $\Upsilon$
decay branching fractions come from the CLEO experiment \cite{ref:cleo}, which placed the 95\% confidence
level upper limit $BF(\Upsilon(3S) \rightarrow \mu^{\pm}\tau^{\mp})<20.3\times 10^{-6}$. This analysis
is more than a factor of four more sensitive to this decay and places the first
upper limits on $BF(\Upsilon(3S) \rightarrow e^{\pm}\tau^{\mp})$. Since the decays we are searching for are 
necessarily mediated by new particles produced 
off-shell in loops, their measurement probes mass scales far exceeding the \pep2\ center-of-mass 
(CM) energy up to the TeV-scale \cite{ref:barbieri}. 

\section{THE \babar\ DETECTOR AND DATASET}
\label{sec:babar}
The data used in this analysis were collected with the \babar\ detector
at the \pep2\ asymmetric-energy \epem\ collider. We search for 
$\Upsilon(3S) \rightarrow e^{\pm}\tau^{\mp}$ and $\Upsilon(3S) \rightarrow \mu^{\pm}\tau^{\mp}$ decays in a sample of $(116.7\pm1.2)\times10^6$ $\Upsilon(3S)$
decays corresponding to an integrated luminosity of 27.5~fb$^{-1}$.  
Data collected at the $\Upsilon(4S)$ corresponding to 77.7~fb$^{-1}$
and data collected 30~MeV below the $\Upsilon(3S)$ resonance corresponding to 2.6~fb$^{-1}$
constitute background control samples which are not expected to contain signal events. An additional data sample collected
at the $\Upsilon(3S)$ corresponding to 1.2~fb$^{-1}$, for which the limit from the CLEO collaboration implies that less than 5 signal events
should be present per channel, is also used as a background control sample which is not included
in the 27.5~fb$^{-1}$ sample.
These control samples are analyzed to verify that a signal yield consistent with zero is obtained. 

Simulated events are also produced and analyzed in order to optimize the selection
and fitting procedure and to compare to data. The background to our events is dominated by continuum QED processes, with
an additional contribution from resonant $\Upsilon(3S) \rightarrow \tau^{+}\tau^{-}$ production. 
The KK2F generator~\cite{ref:kk2f} is used to produce $\mu$-pair and $\tau$-pair events while taking into account the effects of initial state radiation.
The BHWIDE generator~\cite{ref:bhwide} is used to produce Bhabha events, also taking into account initial state radiation. 
The EvtGen generator~\cite{ref:evtgen} is used to produce generic $\Upsilon(3S)$ decays as well as  $1.6\times10^{5}$ signal 
$\Upsilon(3S) \rightarrow e^{\pm}\tau^{\mp}$ and $\Upsilon(3S) \rightarrow \mu^{\pm}\tau^{\mp}$ decays. The efficiency for $q\bar{q}$ events 
($q=u,d,s,c$) and for two-photon processes to pass selection is found to be negligible, so these processes are not included.
The simulated $\mu$-pair, $\tau$-pair and generic $\Upsilon(3S)$ samples correspond to roughly twice the number of events in the $\Upsilon(3S)$ dataset, while
the Bhabha sample, which constitutes a small background to our events, corresponds to roughly half the number of events.
PHOTOS~\cite{ref:photos} is used to simulate radiative corrections, and GEANT~\cite{ref:geant} is used to simulate the interactions of particles traversing the \babar\ detector.

The \babar\ detector is described in detail elsewhere~\cite{ref:babar}.
Charged particle tracking is provided by a five-layer double-sided
silicon vertex tracker (SVT) and a 40-layer drift chamber (DCH). Photons and neutral pions are
identified and their energy deposition is measured using the electromagnetic calorimeter (EMC), which is comprised of 6580 
thallium-doped CsI crystals. These systems are mounted inside a superconducting solenoidal coil providing a 1.5-T
magnetic field. The Instrumented Flux Return (IFR) forms the return yoke of the superconducting coil,
instrumented in the central barrel region with limited streamer tubes for the identification of
muons and the detection of clusters provided by neutral hadrons.

\section{ANALYSIS METHOD}
\label{sec:Analysis}

\subsection{SIGNAL SIGNATURES AND BACKGROUNDS}
\label{subsec:sigmodes}

We search for the decays  $\Upsilon(3S) \rightarrow e^{\pm}\tau^{\mp}$ and $\Upsilon(3S) \rightarrow \mu^{\pm}\tau^{\mp}$.
The signature for these events are two oppositely-charged tracks, a primary electron or muon with CM momentum close to the beam energy,
and a secondary charged lepton or charged pion from a tau decay.  If the tau decays leptonically, we require that the primary and
secondary leptons are of different flavor. If the tau decays hadronically, we require one or two additional neutral pions from
this decay.  Thus we define four signal channels (here and in the following charge conjugate final states are implied):

\begin{itemize}
\item leptonic $e\tau$ channel:~~$\Upsilon(3S) \rightarrow e^{\pm}\tau^{\mp}$, $\tau^{-}\rightarrow \mu^{-}\nu_{\tau}\bar{\nu_{\mu}}$
\item hadronic $e\tau$ channel:~$\Upsilon(3S) \rightarrow e^{\pm}\tau^{\mp}$, $\tau^{-}\rightarrow\pi^{-}\pi^0\nu_{\tau}/\pi^{-}\pi^0\pi^0\nu_{\tau}$
\item leptonic $\mu\tau$ channel:~~$\Upsilon(3S) \rightarrow \mu^{\pm}\tau^{\mp}$, $\tau^{-}\rightarrow e^{-}\nu_{\tau}\bar{\nu_e}$
\item hadronic $\mu\tau$ channel:~$\Upsilon(3S) \rightarrow \mu^{\pm}\tau^{\mp}$,  $\tau^{-}\rightarrow\pi^{-}\pi^0\nu_{\tau}/\pi^{-}\pi^0\pi^0\nu_{\tau}$
\end{itemize}

The $\Upsilon(3S) \rightarrow \ell^{\pm}\tau^{\mp}$, $\tau^{-} \rightarrow \pi^{-}\nu_{\tau}$, 
$\Upsilon(3S) \rightarrow e^{\pm}\tau^{\mp}$, $\tau^{-}\rightarrow e^{-}\nu_{\tau}\bar{\nu_{e}}$ and 
$\Upsilon(3S) \rightarrow \mu^{\pm}\tau^{\mp}$, $\tau^{-}\rightarrow \mu^{-}\nu_{\tau}\bar{\nu_{\mu}}$ decay channels are omitted due to strong
contamination from Bhabha and $\mu$-pair events.
The main source of background to our events comes from $\tau$-pair production, which is dominated by continuum 
production but has a contribution from resonant $\Upsilon(3S)\rightarrow\tau^{+}\tau^{-}$ production as well.   
There is also a background contribution to the $\Upsilon(3S) \rightarrow e^{\pm}\tau^{\mp}$ search from 
Bhabha events in which one of the electrons is misidentified, and to the $\Upsilon(3S) \rightarrow \mu^{\pm}\tau^{\mp}$ search 
from $\mu$-pair events in which one of the muons is misidentified or decays in flight, or an electron is generated in a 
material interaction. 

\subsection{STRATEGY AND SELECTION}
\label{subsec:selection}

Our analysis strategy is to select events with the correct particle types with as high an
efficiency as possible and then fit for the number of events containing an electron or muon with CM momentum close to the beam energy,
which are identified as signal events. A preliminary unblinded analysis is performed using the 1.2~fb$^{-1}$ $\Upsilon(3S)$ data sample 
in order to ensure agreement between data and simulation and to validate the analysis method. A blinded analysis is then performed 
using the full $\Upsilon(3S)$ dataset in which events satisfying $0.95<p_1/E_B<1.00$ are excluded, where $p_1$ is the electron (muon) 
CM momentum for the $\Upsilon(3S) \rightarrow e^{\pm}\tau^{\mp}$ ($\Upsilon(3S) \rightarrow \mu^{\pm}\tau^{\mp}$) search and $E_B$ is 
the beam energy in the CM system, equal to $\sqrt{s}/2$. The blinding criterion rejects more than 99\% of observable 
$\Upsilon(3S) \rightarrow \ell^{\pm}\tau^{\mp}$ decays.

The event selection proceeds in two steps, a preselection for all four signal channels followed by a channel-specific selection.
This procedure is designed to eliminate as much of the Bhabha and $\mu$-pair backgrounds as possible while retaining
high signal efficiency. To pass preselection the event must have two tracks of opposite charge, both consistent with originating 
from the primary interaction point, with opening angle greater than 90$^{\circ}$ in the CM frame. A neutral particle must satisfy 
$E_{DEP} \geq 50$~MeV, where $E_{DEP}$ is the energy deposited in the EMC, and must have a transverse shower profile consistent with that expected from an electromagnetic shower.
The event must satisfy requirements designed to select $e^+e^- \rightarrow \tau^+\tau^-$ events, which are very efficient for our signal.
We require that $\cos(\theta^{lab}_{miss})<$0.9 and $\cos(\theta^{CM}_{miss})>$-0.9, where $\theta^{lab}_{miss}$
($\theta^{CM}_{miss}$) is the polar angle of the missing momentum vector in the lab (CM) frame. These requirements 
suppress events in which particles are lost because they travel along the beam direction; they
also suppress Bhabha, $\mu$-pair and two-photon processes. We require that $M_{VIS}/\sqrt{s}<$0.95, where $M_{VIS}$ is the 
mass of the 4-vector obtained by adding up the 4-vectors of the two tracks plus those of any additional neutral particles
in the event; this requirement also suppresses Bhabha and $\mu$-pair backgrounds.
Finally, we require $(\vec{p}_1+\vec{p}_2)_{\perp}/(\sqrt{s}-|\vec{p}_1|-|\vec{p}_2|)>$0.2, where
$\vec{p}_1$ and $\vec{p}_2$ are the CM momentum vectors of the two tracks. The requirement on this kinematic variable is effective 
at suppressing two-photon processes as well as suppressing beam-gas interactions, in which a particle from the beam undergoes small-angle 
scattering after interacting with residual gas particles in the beam-pipe.

Particle identification is performed using a multivariate algorithm \cite{ref:narsky} that uses measurements from all of the detector
sub-systems, including but not limited to $E/p$ and shower profile from the EMC, the specific ionization $(dE/dx)$ from the tracking detectors,
and the number of hits in the IFR. An electron selector and muon veto, combined with the requirement that the particle falls within the 
angular acceptance of the EMC, are used to identify electrons. A muon selector and electron veto are used to identify muons,
while a charged pion selector, electron veto and muon veto are used to select charged pions. Charged particle misidentification rates are
crucial because they determine the selection efficiencies for Bhabha and $\mu$-pair backgrounds. The probability for a muon to pass the electron
selection criteria is $O(10^{-5})$. The probability for an electron to pass the muon selection criteria, plus the additional
requirement that at least 4 IFR hits are associated to the track as required in the leptonic $e\tau$ channel, is $O(10^{-6})$.
The probability for an electron or muon to pass the charged pion selection criteria is $O(10^{-1})$; this large misidentification
rate is the reason for requiring additional neutral pions in the event.
A pair of photons with invariant mass 0.11~GeV/c$^2<M_{\gamma\gamma}<$0.16~GeV/c$^2$ is selected as a neutral pion candidate.

The channel-specific selection consists of particle identification requirements, requirements on CM momenta, 
and additional kinematic selection criteria aimed at further suppressing the Bhabha and $\mu$-pair backgrounds 
as summarized in Table~\ref{cuts_table}. A summary of estimated signal efficiencies, estimated number of background 
events, and number of data events passing selection for each of the four signal channels 
in the 27.5~fb$^{-1}$ $\Upsilon(3S)$ dataset is displayed in Table~\ref{resultstable}. After including all selection requirements, 
typical signal efficiencies are 4-6\% (including the $\tau$ branching fraction), and typical $\tau$-pair background efficiencies 
are $(6-10)\times10^{-4}$, depending on the signal channel.

Due to the combination of particle identification selectors and vetoes used in this analysis, it is in general not possible
for a single event to pass selection for more than one signal channel. The exception is that it is possible for a $\tau$-pair event decaying
to a final state containing an electron and a muon with $p_e/E_B$, $p_{\mu}/E_B>0.75$ to pass selection for both the leptonic $e\tau$ and
leptonic $\mu\tau$ signal channels. The probability for an event passing selection for one of the leptonic channels to pass selection
for the other leptonic channel is determined to be 2\%, in which case the event is included
in both leptonic signal channels.

\begin{table}
\begin{center}
\caption{ Channel-specific selection for the four signal channels. The subscript 1
refers to the primary track which is the electron (muon) for the $\Upsilon(3S) \rightarrow e^{\pm}\tau^{\mp}$ 
($\Upsilon(3S) \rightarrow \mu^{\pm}\tau^{\mp}$) search; the subscript 2 refers to the other 
track. CM momenta are denoted by $p$, $E_B$ is the beam energy and $\Delta\phi^{CM}$ is
the difference between the azimuthal angles of the two 
tracks in the CM frame. The secondary track CM transverse momentum is denoted by $p_{T2}$ , and 
$N_{IFR2}$ is the number of IFR hits associated to the secondary track. 
The invariant mass of the $\pi^{\pm}\pi^{0}$ system is denoted by M($\pi^{\pm}\pi^{0}$). If there are two 
neutral pions in the event, the $\pi^0$ giving the $\pi^{\pm}\pi^{0}$ mass closest to $m_{\rho}$=0.77~GeV/c$^2$ is 
chosen. The requirement on M($\pi^{\pm}\pi^{0}\pi^{0}$) is included only if there are two neutral 
pions in the event. Empty table entries indicate that no cut is used for the given channel.}
\begin{tabular}{|l|c|c|c|c|}
\hline
Quantity  & leptonic $e\tau$ & hadronic $e\tau$ & leptonic $\mu\tau$ & hadronic $\mu\tau$  \\
\hline
PID & 1e, 1$\mu$ &  1$e$, 1$\pi^{\pm}$, 1 or 2 $\pi^0$'s &
1$e$, 1$\mu$ & 1$\mu$, 1$\pi^{\pm}$, 1 or 2 $\pi^0$'s \\
\hline
$p_1/E_B$ & $>$0.75 & $>$0.75 & $>$0.75 & $>$0.75 \\
\hline
$p_{T2}/E_B$ &  & $>$0.05  &  & \\
\hline
$p_2/E_B$ & & $<$0.8 & & $<$0.8 \\
\hline
$\Delta\phi^{CM}$ & &  & $<$172$^{\circ}$ & \\
\hline
$N_{IFR2}$ & $>$3 & & &  \\
\hline
M($\pi^{\pm}\pi^{0}$) & & 0.4-1.1~GeV/c$^2$ & & 0.4-1.1~GeV/c$^2$ \\
\hline
M($\pi^{\pm}\pi^{0}\pi^{0}$) & & 0.6-1.5~GeV/c$^2$ & & 0.6-1.5~GeV/c$^2$ \\
\hline
\end{tabular}
\label{cuts_table}
\end{center}
\end{table}

\begin{table}
\caption{
Summary of selection for 27.5~fb$^{-1}$ of $\Upsilon(3S)$ data and simulated events.
Shown are estimated signal efficiencies and estimated number of background events, determined from the simulated
event samples, and number of data events passing selection in the four signal channels. The signal efficiencies include the $\tau$ branching fractions.
All errors are statistical only and the errors in the $\tau$ branching fractions are included in the signal
efficiency uncertainties. Systematic effects leading to potential discrepancies between the number of data and simulated events passing selection
are discussed in Sec.~\ref{sec:Systematics}}
\begin{center}
\begin{tabular}{|l|c|c|c|c|c|}
\hline
Quantity & leptonic $e\tau$ & hadronic $e\tau$ & leptonic $\mu\tau$ & hadronic $\mu\tau$\\
\hline
$\epsilon_{SIG}$ (\%) & 4.72$\pm$0.05  & 4.94$\pm$0.06  & 4.16$\pm$0.05 & 6.21$\pm$0.06 \\
\hline
$N_{BKG}$  & 18966$\pm$100  &  20524$\pm$101 & 19995$\pm$100 &  28087$\pm$119 \\
\hline
$N_{DATA}$ & 18720  & 20548 & 19966  & 27479 \\
\hline
\end{tabular}
\label{resultstable}
\end{center}
\end{table}

\begin{figure}[!htb]
\begin{center}
\includegraphics[width=1 \textwidth]{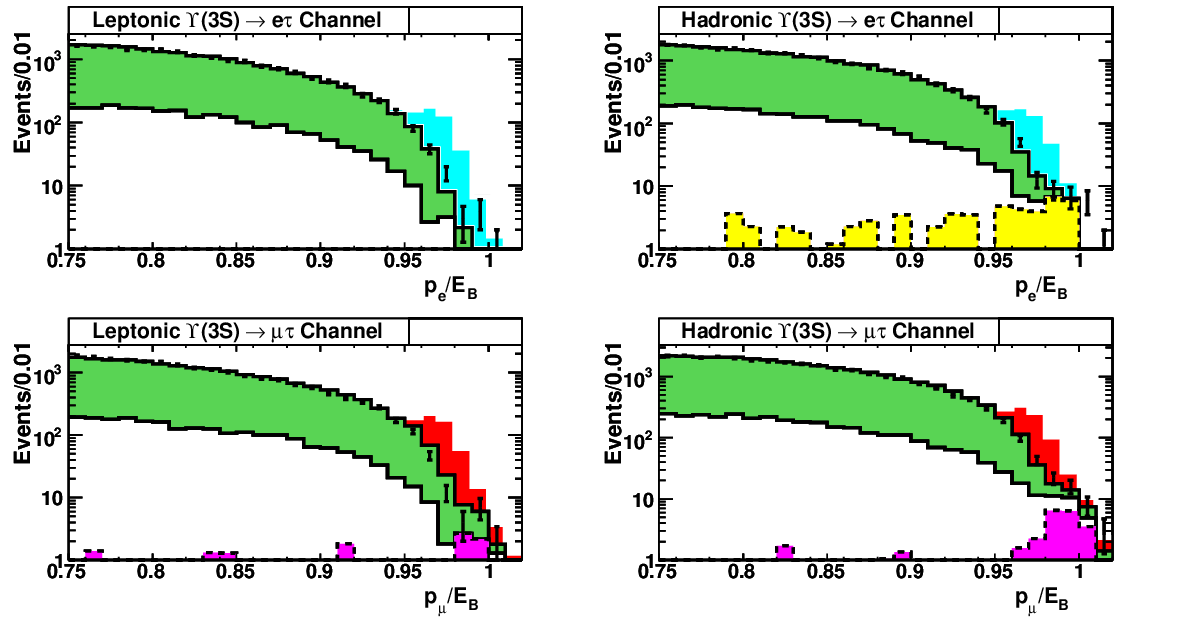}
\caption{
The beam-energy-normalized primary track CM momentum distributions $x$=$p_1/E_B$ for the
selected data and simulated events in the four signal channels.  Data are denoted by points with 
statistical error bars, Standard Model background processes are denoted by histograms with solid
or dotted outlines, and signal events are denoted by histograms without solid outlines, where
$BF(\Upsilon(3S) \rightarrow \ell^{\pm}\tau^{\mp})$ has been set to 10$^{-4}$. Solid green is 
continuum $\tau$-pair production and the unshaded region is resonant $\tau$-pair production. For the 
$e\tau$ channels, the yellow histogram with a dotted outline is Bhabha events and the solid light blue 
histogram with no outline is signal $\Upsilon(3S) \rightarrow e^{\pm}\tau^{\mp}$ production. 
For the $\mu\tau$ channels, the magenta histogram with a dotted outline is $\mu$-pair events and the solid red 
histogram with no outline is signal $\Upsilon(3S) \rightarrow \mu^{\pm}\tau^{\mp}$ production.}

\label{fig1}
\end{center}
\end{figure}

\subsection{FIT METHOD}
\label{subsec:fitmethod}

The discriminant kinematic variable separating signal from background is the beam-energy-normalized
primary track CM momentum $x$=$p_1/E_B$, where the primary track is defined to be the
electron in the $\Upsilon(3S) \rightarrow e^{\pm}\tau^{\mp}$ search and the muon in the $\Upsilon(3S) \rightarrow \mu^{\pm}\tau^{\mp}$ search.
A comparison of the $x$ distributions between selected data and simulated events is displayed in Fig.~\ref{fig1}.
After selection an unbinned, extended maximum likelihood fit is performed. A probability density function (PDF) consisting of a sum of three components is fit to the
measured $x$ distribution for each signal channel. The three components are signal $\Upsilon(3S) \rightarrow \ell^{\pm}\tau^{\mp}$ production, $\tau$-pair background, 
and Bhabha ($\mu$-pair) background for the $\Upsilon(3S) \rightarrow e^{\pm}\tau^{\mp}$ ($\Upsilon(3S) \rightarrow \mu^{\pm}\tau^{\mp}$) search, with the yields of the three components
left as free parameters in the fit. PDFs for signal, Bhabha, and $\mu$-pair backgrounds are determined using $x$ distributions from simulated events
which are required to pass the same selection criteria as data. The procedure for determining the $\tau$-pair
background PDF is discussed below.  The signal yield $N_{SIG}$ is extracted by the fit and is used to measure the corresponding branching fraction 
according to $BF=N_{SIG}/(\epsilon_{SIG}\times N_{\Upsilon(3S)})$, where $\epsilon_{SIG}$ is the signal selection efficiency and $N_{\Upsilon(3S)}$ 
is the number of produced $\Upsilon(3S)$ decays.

For $\Upsilon(3S) \rightarrow \ell^{\pm}\tau^{\mp}$ decays the $x$ distribution is sharply peaked at $x$=0.97 because some of the collision energy is 
carried away by the $\tau$ mass. The $x$ distribution for the signal has a width of about 0.01 and a radiative tail,
which is more pronounced for the $e\tau$ than for the $\mu\tau$ decays. The resulting $x$ distribution is well-described by a
modified Crystal Ball function \cite{ref:cb}, hereafter referred to as a double-sided Crystal Ball function, which has both a low-energy and a high-energy tail.
The shape of this function is extracted from fits to the simulated $\Upsilon(3S) \rightarrow \ell^{\pm}\tau^{\mp}$
$x$ distributions. The Bhabha and $\mu$-pair backgrounds have a peaking Gaussian component near $x$=1 (approximately three times the resolution
above the signal peak) also with a width of about 0.01, and a broad component modeled by an Argus function \cite{ref:argus} which truncates near $x$=1. 
The shape of the Argus plus Gaussian fit function is extracted from fits to the $x$ distributions of simulated Bhabha and $\mu$-pair events.

\begin{figure}[!t]
\begin{center}
\includegraphics[width=1 \textwidth]{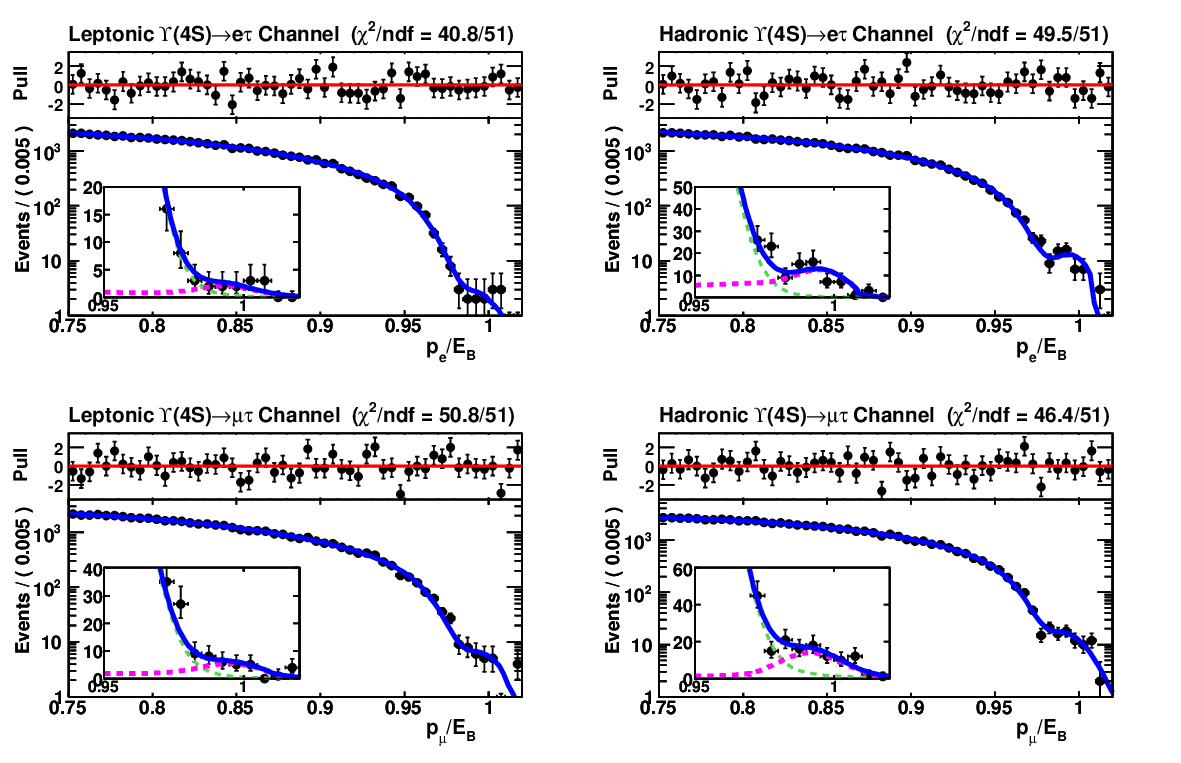}
\caption{Fit results for the 77.7~fb$^{-1}$ $\Upsilon(4S)$ dataset for the four signal channels. The dashed
lines indicate the two component PDFs, the solid line indicates their sum. The thin green dashed line
is the $\tau$-pair background PDF, the thick magenta dashed line is the Bhahba background PDF for the $e\tau$
channels and the $\mu$-pair background PDF for the $\mu\tau$ channels. The inset shows a close-up of the region 0.95$<x<$1.02.}
\label{fig2}
\end{center}
\end{figure}

The PDF describing the $\tau$-pair background $x$ distribution is given by the convolution of a polynomial and
a detector resolution function. The polynomial is given by Eq.~\ref{poly}:
\begin{equation}
p(x)=(1-x/x_{MAX})+c_2(1-x/x_{MAX})^2+c_3(1-x/x_{MAX})^3
\label{poly}
\end{equation}
in which $x_{MAX}$ determines the kinematic cut-off and $c_2$ and $c_3$ determine the polynomial shape. 
The detector resolution function is extracted from fits to lepton momentum resolution distributions from simulated $\tau$-pair
events. The resulting distributions are found to be well-described
by a double-sided Crystal Ball function, whose width varies linearly as a function of generated momentum and is extrapolated to
$0.96\times E_B$, approximately 1$\sigma$ below the $\tau$-pair kinematic cut-off. The value of $x_{MAX}$ is extracted from a fit to the 
77.7~fb$^{-1}$ $\Upsilon(4S)$ data control sample as shown in Fig.~\ref{fig2}, while the polynomial shape parameters, 
which are not strongly correlated with the signal yield, are left as free parameters in the fit to $\Upsilon(3S)$ data.

\subsection{FIT VALIDATION}
\label{subsec:fitvalidation}

To validate the fit procedure, we perform fits to the $\Upsilon(4S)$ data control sample in order to verify
that a signal yield consistent with zero is obtained for each of the four signal channels. The efficiency for $B^0\bar{B^0}/B^+B^-$ events to pass 
selection is negligible, so that the event compositions for the $\Upsilon(4S)$ and $\Upsilon(3S)$ data samples are similar, with the exception that the $\Upsilon(3S)$ data sample contains an $O(10\%)$ contribution from resonant $\tau$-pair production. The $\Upsilon(4S)$ dataset is divided into a 2/3 (51.8~fb$^{-1}$)
`control' sample and a 1/3 (25.9~fb$^{-1}$) `fit' sample, where the size of the fit sample is chosen to be comparable to the full
$\Upsilon(3S)$ dataset. The polynomial shape and cut-off are extracted from a fit to the `control' sample, in which the
$\tau$-pair and Bhabha/$\mu$-pair background yields are floated but the signal yield is fixed to zero. The fit is then repeated using the
`fit' sample, in which the polynomial shape and cut-off are fixed but the yields for signal, $\tau$-pair, and Bhabha/$\mu$-pair backgrounds
are floated. Signal yields consistent with zero within $\pm$1.4$\sigma$ are obtained for the four signal channels. 
We perform similar studies using the 2.6~fb$^{-1}$ $\Upsilon(3S)$ off-resonance sample and the 1.2~fb$^{-1}$ $\Upsilon(3S)$ on-resonance sample,
for which the limit from the CLEO collaboration implies that less than 5 events should be present per channel. These studies confirm the results of the
fit validation study using the $\Upsilon(4S)$ data control sample.

A large number of pseudo-experiments are also performed in order to further validate the fit procedure. Sample data sets are generated from the
background PDF, simulated signal events are added and the fit is performed using the resulting simulated dataset. 
This procedure is repeated multiple times and the extracted signal yield, error and pull are recorded.  This study confirms that 
no large biases are observed and that the calculated statistical errors are accurate. 

\section{SYSTEMATIC ERROR STUDIES}
\label{sec:Systematics}

The decay branching fractions are determined using the extracted signal yield, estimated signal efficiency, and number of produced
$\Upsilon(3S)$ decays. The uncertainties in these quantities are summarized in Table~\ref{syst_table}.
There is also a systematic uncertainty arising from potential bias in the fit procedure. 

\begin{table}
\caption{Summary of systematic uncertainties in the signal yield, signal efficiency and number of produced $\Upsilon(3S)$ decays.
The uncertainties in the signal efficiencies and number of $\Upsilon(3S)$ decays include both statistical and systematic effects. 
Also shown is the uncertainty due to potential bias in the fit procedure.}
\begin{center}
\begin{tabular}{|l|c|c|c|c|}
\hline
Quantity & leptonic $e\tau$ & hadronic $e\tau$ & leptonic $\mu\tau$ & hadronic $\mu\tau$ \\ 
\hline
$N_{SIG}$ & 5.8 events & 8.2 events & 6.7 events & 11.5 events \\
\hline
$\epsilon_{SIG}$ & 2.0\% & 3.2\% & 2.2\% & 2.9\% \\
\hline
$N_{\Upsilon(3S)}$  & 1.0\% & 1.0\% & 1.0\% & 1.0\% \\
\hline
Fit Bias  & 1.6 events & 1.3 events & 0.8 events & 1.3 events \\
\hline
\end{tabular}
\label{syst_table}
\end{center}
\end{table}

The dominant systematic uncertainty in the decay branching fractions comes from the
uncertainty in the extracted signal yields due to uncertainties in the PDF shape parameters. To assess these uncertainties
each parameter $p_i$ is varied by its uncertainty and the resulting change in signal yield $\delta_i$=$\Delta N_{SIG}$
is determined. The total systematic uncertainty in the signal yield is given by $\delta_{TOT}=\sqrt{\vec{\delta}^T C \vec{\delta}}$,
where $\vec{\delta}=<\delta_1 \cdot \cdot \cdot \delta_N>$ and C is the parameter correlation matrix, giving a systematic uncertainty 
in the signal yield of 6-12 events depending on the signal channel.

The signal efficiency is determined using simulated events, so there is a systematic uncertainty arising from any potential discrepancies between 
data and simulation. To assess this uncertainty, the relative difference between the yields for data
and simulated events is taken using $\tau$-pair control samples from a portion of the sideband of the $x$ distribution defined by $0.8<x<0.9$. 
This procedure gives systematic uncertainties in the signal efficiency of 2-3\% depending
on the signal channel, due to uncertainties in the particle identification, tracking, trigger and kinematic selection 
efficiencies.  The statistical uncertainty in the signal efficiency is about 1\% for each channel.  
The number of produced $\Upsilon(3S)$ decays in the dataset is estimated to be $N_{\Upsilon(3S)}=(116.7\pm1.2)\times10^6$ by counting
the number of multihadron events in the dataset, giving a 1\% systematic uncertainty.

To assess the potential bias in the fit procedure, the study described in the previous section using a large number of pseudo-experiments
is repeated with the generated signal yield fixed to the value extracted by the fit to 27.5~fb$^{-1}$ $\Upsilon(3S)$ data. For those signal channels
in which a negative signal yield is extracted, the generated signal yield is fixed to zero. The fit is performed multiple times
and the deviation from zero in the distribution of extracted minus generated signal yield is taken as the uncertainty due to the fit 
bias, yielding uncertainties of about 1 event count.

\section{RESULTS}
\label{sec:Results}

\begin{figure}[!htb]
\begin{center}
\includegraphics[width=1 \textwidth]{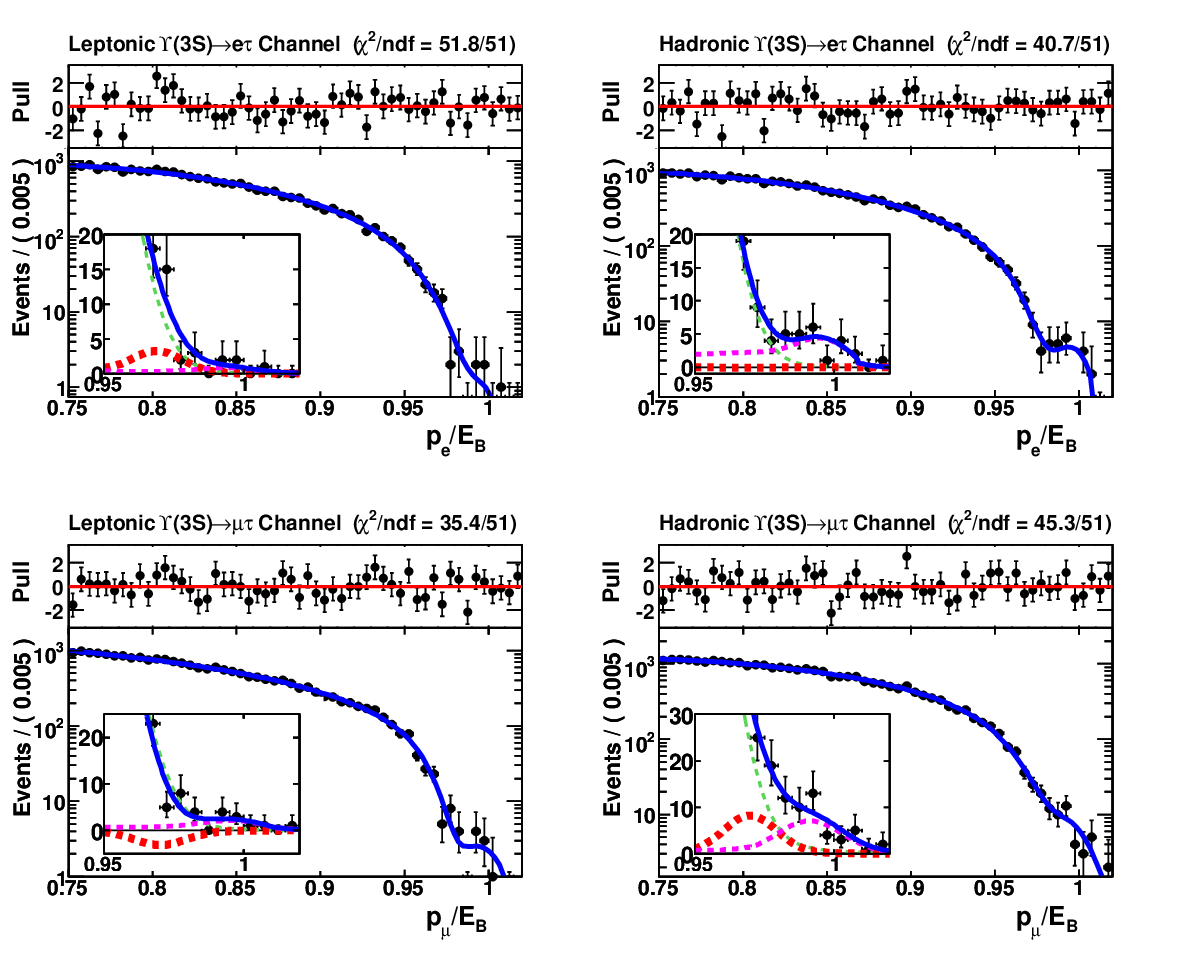}
\caption{Fit results for the 27.5~fb$^{-1}$ $\Upsilon(3S)$ dataset for the four signal channels. The dashed
lines indicate the three component PDFs, the solid line indicates their sum. The thin green dashed line
is the $\tau$-pair background PDF, the medium magenta dashed line is the Bhahba background PDF for the $e\tau$
channels and the $\mu$-pair background PDF for the $\mu\tau$ channels, and the
thick red dashed line is the signal PDF. The inset shows a close-up of the region 0.95$<x<$1.02.}
\label{fig3}
\end{center}
\end{figure}

\begin{figure}[!htb]
\begin{center}
\includegraphics[width=1 \textwidth]{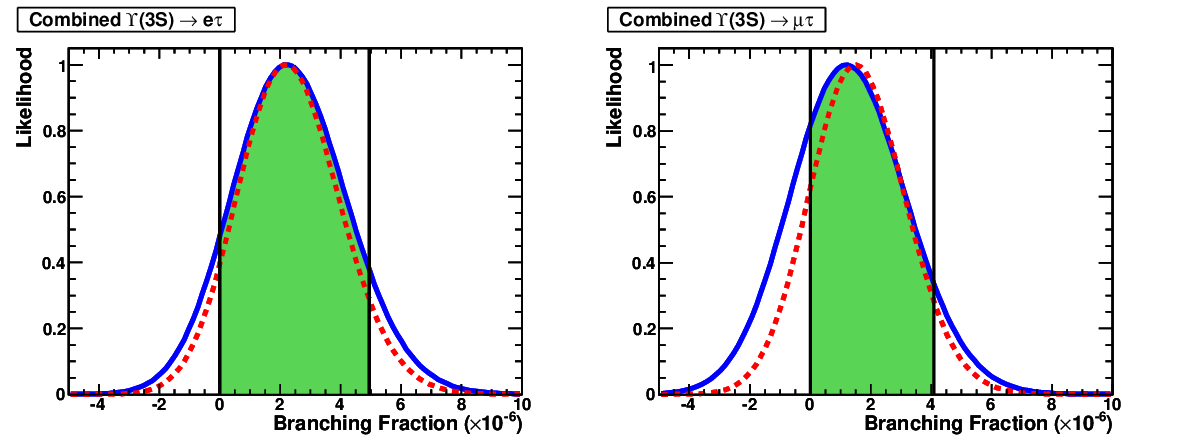}
\caption{The results of the likelihood scan. The likelihood curves for the two 
$\Upsilon(3S) \rightarrow e^{\pm}\tau^{\mp}$ and 
two $\Upsilon(3S) \rightarrow \mu^{\pm}\tau^{\mp}$ signal
channels have been multiplied to obtain the combined likelihood curves. The red dashed lines indicate statistical 
uncertainties only, the solid blue lines have systematic uncertainties incorporated. The vertical lines bound 90\%
of the positive integral of the total likelihood curve,  which is shaded in green.}
\label{fig4}
\end{center}
\end{figure}

The fit results for the full $\Upsilon(3S)$ dataset are displayed in Table~\ref{yieldtable} and in Fig.~\ref{fig3}. After including statistical and systematic uncertainties, 
the extracted signal yields are all consistent with zero within $\pm$2.1$\sigma$. The upper limits, mean values, and asymmetric errors on the decay branching fractions are
determined by performing a likelihood scan.  For each of the four signal channels,
the signal yield is scanned in steps from -50 to 150, the fit is performed with the signal yield fixed at that value, and the
likelihood $L$ is extracted. The signal yield is converted to a branching fraction according to $BF=N_{SIG}/(\epsilon_{SIG} \times N_{\Upsilon(3S)})$,
giving the likelihood as a function of branching fraction.  Systematic effects which are uncorrelated between the signal channels
are included in the likelihood curves for each channel individually, resulting in widening of the likelihood function. The two 
$e\tau$ ($\mu\tau$) likelihood curves are multiplied to obtain the combined $e\tau$ ($\mu\tau$) likelihood curve, and correlated systematic effects are then included. The branching fraction upper limit at 90\% confidence level is determined by integrating 
the likelihood function and finding UL such that $\int_0^{UL} L\,d(BF)$/$\int_0^{\infty} L\,d(BF)$=0.9, which is equivalent to a Bayesian 
upper limit extraction in which the prior is taken to be the step function $\theta(0)$. The most probable value is taken to be the
BF corresponding to the maximum of the likelihood function, and the asymmetric errors are determined by finding $\pm\Delta$(BF) such that
$\Delta(-2 \log(L)$)=1. The likelihood scan is displayed in Fig.~\ref{fig4}, and the results of the scan are summarized in 
Table~\ref{finalresultstable}.

\begin{table}
\caption{Summary of the fit results for the four signal channels of  the $\Upsilon(3S)$ dataset. 
Shown are the extracted signal yield $N_{SIG}$, the extracted Bhabha/$\mu$-pair background yield $N_{PKBKG}$, and
the extracted $\tau$-pair background yield $N_{BKG}$. The first error is the parabolic statistical error, the second error (displayed
only for the signal yield) is the systematic error from the PDF shape uncertainties and the fit bias.
The signal yields for all channels are consistent with zero within $\pm2.1\sigma$ after including
statistical and systematic uncertainties.}
\begin{center}
\begin{tabular}{|l|l|l|l|l|}
\hline
Yield  & leptonic $e\tau$ & hadronic $e\tau$ & leptonic $\mu\tau$ & hadronic $\mu\tau$ \\ 
\hline
$N_{SIG}$ & $21 \pm 12 \pm 6$ & $-1 \pm 14 \pm 8$ & $-16 \pm 9 \pm 7$ & $42 \pm 17 \pm 12$ \\
\hline
$N_{PKBKG}$ & $25 \pm 15$ & $63 \pm 16$ & $35 \pm 13$ & $57 \pm 12$ \\
\hline
$N_{BKG}$ & $19611 \pm 141$ & $21523 \pm 147$ & $20923 \pm 145$ & $28661 \pm 170$ \\
\hline
\end{tabular}
\label{yieldtable}
\end{center}
\end{table}

\begin{table}
\caption{Summary of the results of the likelihood scan. Displayed are the 90\% confidence level upper limits
and most probable values (MPV), with negative and positive asymmetric errors included. }
\begin{center}
\begin{tabular}{|l|c|c|}
\hline
& UL & MPV  \\
\hline
BF($\Upsilon(3S) \rightarrow e^{\pm}\tau^{\mp}$) ($\times 10^{-6}$) & $<5.0$ & $2.2_{-1.8}^{+1.9}$ \\
\hline
BF($\Upsilon(3S) \rightarrow \mu^{\pm}\tau^{\mp}$) ($\times 10^{-6}$) & $<4.1$ & $1.2_{-1.9}^{+1.9}$ \\
\hline
\end{tabular}
\label{finalresultstable}
\end{center}
\end{table}

\section{CONCLUSIONS}
\label{sec:Conclusions}

This paper presents a search for the CLFV decays $\Upsilon(3S)\rightarrow e^{\pm}\tau^{\mp}$ and
$\Upsilon(3S)\rightarrow \mu^{\pm}\tau^{\mp}$. A maximum likelihood fit is performed using the primary lepton 
CM momentum distribution to extract the signal yield. No statistically significant signal is observed
and the results are used to place the following 90\% confidence level upper limits on the decay
branching fractions: $BF(\Upsilon(3S) \rightarrow e^{\pm}\tau^{\mp})<5.0\times 10^{-6}$ and $BF(\Upsilon(3S) \rightarrow \mu^{\pm}\tau^{\mp})<4.1\times 10^{-6}$.
These results represent the first upper limits on $BF(\Upsilon(3S) \rightarrow e^{\pm}\tau^{\mp})$ and represent a sensitivity improvement
of more than a factor of four with respect to the previous upper limit on $BF(\Upsilon(3S) \rightarrow \mu^{\pm}\tau^{\mp})$~\cite{ref:cleo}.

Effective field theory allows the effects of BSM physics at some large mass scale contributing to CLFV $\Upsilon(3S)$ decays to be parameterized at low energy 
by a four-fermion interaction with coupling constant $\alpha_N$ and mass scale $\Lambda$. This allows the following relation to be derived \cite{ref:silagadze,ref:black}:
\begin{equation}
\frac{\Gamma(\Upsilon(3S) \rightarrow \ell^{\pm}\tau^{\mp})} {\Gamma(\Upsilon(3S) \rightarrow \ell^{+}\ell^{-})}=\frac{1}{2 q_b^2}  \left(\frac{\alpha_N^{(\ell\tau)}}{\alpha}\right)^2 \left({\frac{M_{\Upsilon(3S)}}{\Lambda^{(\ell\tau)}}}\right)^{4} (\ell=e,\mu)
\label{lfvequation}
\end{equation}
in which $q_b$ is the charge of the $b$ quark. The dilepton branching fraction of the $\Upsilon(3S)$ is taken to be the average of the PDG~\cite{ref:pdg} values
of $BF(\Upsilon(3S)\rightarrow e^+e^-)$ and $BF(\Upsilon(3S)\rightarrow \mu^+\mu^-)$, giving $BF(\Upsilon(3S)\rightarrow \ell^+\ell^-)=(2.18 \pm 0.15)\times 10^{-2}$.
Using this result, assuming strong vector coupling with $\alpha_N^{(e\tau)}$=$\alpha_N^{(\mu\tau)}$=1, and taking into account the uncertainty in the 
dilepton branching fractions of the $\Upsilon(3S)$, our results place the 90\% confidence level lower limits $\Lambda^{(e\tau)}>$1.4~TeV and 
$\Lambda^{(\mu\tau)}>$1.5~TeV on the mass scale of BSM physics contributing to CLFV in the bottomonium sector. This result improves upon the previous result from the CLEO collaboration~\cite{ref:cleo}, which used a similar analysis technique to place the 95\% confidence level lower limit $\Lambda^{(\mu\tau)}>$1.34~TeV.

\section{ACKNOWLEDGMENTS}
\label{sec:Acknowledgments}

We are grateful for the 
extraordinary contributions of our \pep2\ colleagues in
achieving the excellent luminosity and machine conditions
that have made this work possible.
The success of this project also relies critically on the 
expertise and dedication of the computing organizations that 
support \babar.
The collaborating institutions wish to thank 
SLAC for its support and the kind hospitality extended to them. 
This work is supported by the
US Department of Energy
and National Science Foundation, the
Natural Sciences and Engineering Research Council (Canada),
the Commissariat \`a l'Energie Atomique and
Institut National de Physique Nucl\'eaire et de Physique des Particules
(France), the
Bundesministerium f\"ur Bildung und Forschung and
Deutsche Forschungsgemeinschaft
(Germany), the
Istituto Nazionale di Fisica Nucleare (Italy),
the Foundation for Fundamental Research on Matter (The Netherlands),
the Research Council of Norway, the
Ministry of Education and Science of the Russian Federation, 
Ministerio de Educaci\'on y Ciencia (Spain), and the
Science and Technology Facilities Council (United Kingdom).
Individuals have received support from 
the Marie-Curie IEF program (European Union) and
the A. P. Sloan Foundation.

\end{document}